\documentclass[conference]{IEEEtran}
\IEEEoverridecommandlockouts
\usepackage{cite}
\usepackage{amsmath,amssymb,amsfonts}
\usepackage{algorithmic}
\usepackage{graphicx}
\usepackage{textcomp}
\usepackage{xcolor}
\usepackage[hyphens]{url}

\usepackage{pgfplots}
\pgfplotsset{compat=newest}

\usepackage{amsfonts}
\usepackage{colortbl}
\usepackage{xspace}
\usepackage{lipsum}

\usepackage{fontawesome}

\usepackage{caption}
\usepackage{subcaption}

\usepackage[]{hyperref}

\usepackage{enumitem}

\usepackage[export]{adjustbox}

\usepackage{lineno}

\usepackage{xcolor}
\usepackage{soul}        %

\colorlet{pink}{red!40}
\colorlet{cyanblue}{cyan!80}

\definecolor{myred}{RGB}{250,214,221}
\definecolor{myblue}{RGB}{190,228,254}

\usepackage{tikz}
\usetikzlibrary{calc}

\usepackage{pgf}

\newdimen\mydim

\newcommand{\gettikzxy}[3]{%
  \tikz@scan@one@point\pgfutil@firstofone#1\relax
  \edef#2{\the\pgf@x}%
  \edef#3{\the\pgf@y}%
}

\newdimen\XCoord
\newdimen\YCoord

\usepackage{mathtools}
\usepackage{bm}
\usepackage{empheq}
\usepackage{amsthm}

\usepackage{resizegather} %

\usepackage[linguistics]{forest}

\usepackage[short]{optidef}

\usepackage{siunitx}
\newcommand{\unsim}{\mathord{\sim}}

\usepackage[ruled,vlined,linesnumbered,resetcount,algosection,nofillcomment]{algorithm2e}

\SetKwInput{KwData}{Input}
\SetKwInput{KwResult}{Output}
\SetKwFor{Pfor}{parallel for}{}{}
\SetKwFor{PipFor}{pipelined for}{}{}
\SetKwFor{PipForEach}{pipelined foreach}{do}{endfch}
\SetKw{KwBy}{by}
\SetKw{KwNot}{not}
\SetKw{KwBreak}{break}
\SetKwFor{Loop}{loop}{}{}

\SetKwRepeat{Do}{do}{while}
\SetKwRepeat{PipDo}{pipelined do}{while}
\SetKwProg{Pn}{function}{:}{\KwRet}
\SetKw{Output}{output}

\let\oldnl\nl%
\newcommand{\nonl}{\renewcommand{\nl}{\let\nl\oldnl}}%

\definecolor{mycommentcolor}{rgb}{0.0, 0.5, 0.6}

\SetCommentSty{mycommfont}

\usepackage{comment}

\usepackage[normalem]{ulem}

\newcommand{\myrevision}{false}
\newcommand{\myrevisionbigpage}{false}

\usepackage{ifthen}

\ifthenelse{\equal{\myrevisionbigpage}{true}}
{
  \paperwidth=\dimexpr \paperwidth + 8cm\relax
  \oddsidemargin=\dimexpr\oddsidemargin + 4cm\relax
  \evensidemargin=\dimexpr\evensidemargin + 4cm\relax
  \marginparwidth=\dimexpr \marginparwidth + 4cm\relax

  \paperheight=\dimexpr \paperheight + 8cm\relax
}
{
}

\ifthenelse{\equal{\myrevision}{true}}
  {%
\usepackage[colorinlistoftodos]{todonotes}
\setuptodonotes{fancyline, color=blue!30, size=\small}

    \newcommand\td[1]{\todo{#1}}
    \newcommand\tds[1]{\todo{\sout{#1}}}
    \newcommand\suggestion[2]{\todo[backgroundcolor=green!20!white]{ #2}}
    \newcommand\solvedsuggestion[2]{\todo[backgroundcolor=green!20!white]{\sout{#2}}}
    
    \newcommand\rv[1]{{\color{blue} {#1}}}

  }
  {%
    \newcommand\td[1]{}
    \newcommand\tds[1]{}
    \newcommand\suggestion[2]{}
    \newcommand\solvedsuggestion[2]{}
    
    \newcommand\rv[1]{{#1}}
    
  }

\newcommand{\setword}[2]{%
  \phantomsection
  #1\def\@currentlabel{\unexpanded{#1}}\label{#2}%
}

\usepackage{etoolbox}  %

\usepackage[savepos,totpages,titleref,dotfill,counter,user]{zref}
\usepackage{pageslts}
\usepackage{atbegshi}     %
\usepackage{xfp}  %

\makeatletter

\newcommand{\refacmlineno}[1]{%
  \edef\Ysp{\zposy{acmlinenostart:#1}}%
  \edef\Xsp{\zposx{acmlinenostart:#1}}%
  \edef\Psp{\zref@extract{zlabel:acmlinenostart:#1}{abspage}}%
  \edef\Yep{\zposy{acmlinenoend:#1}}%
  \edef\Xep{\zposx{acmlinenoend:#1}}%
  \edef\Pep{\zref@extract{zlabel:acmlinenoend:#1}{abspage}}%
  \hyperref[acmline:#1]{
  {\textcolor{red}{Line $\fpeval{ (\Psp - 2) * 58 * 2 + (round((\Xsp + 10457825) / 20915650) - 1) * 58  +  round((580 + 121 - round((\Ysp / 65536) )) /11) + 1  }\unsim \fpeval{ (\Pep - 2) * 58 * 2 + (round((\Xep + 10457825) / 20915650) - 1) * 58 +  round((580 + 121 - round((\Yep / 65536) )) /11) + 1}$}%
  }%
  }%
}

\makeatother

\usepackage{booktabs}
\usepackage{tabularx}
\usepackage{multirow,multicol,threeparttable, tablefootnote}
\usepackage{makecell}
\newcolumntype{Y}{>{\centering\arraybackslash}X}

\usepackage{mdframed}
\mdfdefinestyle{myframe}{%
    skipabove=0pt,
    skipbelow=0pt
}

\usepackage[most]{tcolorbox}

\usepackage{cleveref}
\newcommand{\mref}[2]{\textcolor{red}{\hyperref[#1]{#2}}}
\crefname{algocf}{alg.}{algs.}
\Crefname{algocf}{Algorithm}{Algorithms}

\usepackage{pifont}

\pgfmathsetmacro{\nodebasesize}{1} %
\pgfmathsetmacro{\nodeinnersep}{0.05}

\definecolor{softblue}{HTML}{31859B}
\definecolor{softred}{HTML}{923931}

\def\BibTeX{{\rm B\kern-.05em{\sc i\kern-.025em b}\kern-.08em
    T\kern-.1667em\lower.7ex\hbox{E}\kern-.125emX}}
\begin{document}

\pdfpagewidth=8.5in
\pdfpageheight=11in

\newcommand{\iscasubmissionnumber}{1323}

\pagenumbering{arabic}

\title{Approaching Shannon Bound with Lossless LLM Weight Compression}

\author{
\IEEEauthorblockN{
    Hongshi Tan\textsuperscript{*},
    Yao Chen\textsuperscript{\dag},
    Gustavo Alonso\textsuperscript{\S},
    Weng-Fai Wong\textsuperscript{*},
    Bingsheng He\textsuperscript{*}
}
\IEEEauthorblockA{
    \textsuperscript{*}School of Computing, National University of Singapore, Singapore \\
    \textsuperscript{\dag}School of Computer Science and Technology, Huazhong University of Science and Technology, China \\
    \textsuperscript{\S}Department of Computer Science, ETH Zurich, Switzerland\\
    hongshi@u.nus.edu,  chenyao\_cs@hust.edu.cn, alonso@inf.ethz.ch, \{dcswwf, dcsheb\}@nus.edu.sg
}
}

\maketitle
\thispagestyle{plain}
\pagestyle{plain}

\begin{abstract}

Large language models (LLMs) now scale to trillions of parameters, driving weight storage into the terabyte regime and creating an acute mismatch with GPU memory capacity. Although lossless compression is widely effective in other domains, it remains underutilized in LLM systems. Through a comprehensive entropy study across models from 1.5B to 405B parameters and numeric formats ranging from bf16 to int4 and AWQ/SQ8, we find that LLM weights contain far less intrinsic randomness than their stored bitwidth implies, their effective entropy is \textbf{2--10$\times$ lower}, indicating that \textbf{up to a 10$\times$ footprint reduction} is theoretically achievable without altering any weight values. Leveraging this insight, we introduce a \textbf{tile-level, on-the-fly lossless decompression framework} based on Asymmetric Numeral Systems that aligns decoding with the GEMM tiling pattern of GPU inference. Our design achieves \textbf{bit-rates within 0.01--0.1 bits of the Shannon limit} \rv{across a wide range of LLM numerical formats}, demonstrating that nearly all statistical redundancy is eliminated.
\rv{Integrated into the SGLang serving framework with multi-GPU support, our approach increases the maximum batch size of Qwen-14B from 47 to 75, improving throughput by up to 1.2$\times$. On Mixtral-176B, the feasible batch size increases from 20 to 95 (4.8$\times$), yielding up to 1.6$\times$ throughput improvement. Compared to state-of-the-art lossless compression approaches NeuZip and DFloat11, our design further improves throughput by up to 11$\times$.}

\end{abstract}

\section{Introduction}
\label{sec:introduction}

Large language models (LLMs) have become the foundational backbone of modern AI systems~\cite{liFundamentalCapabilitiesLarge2024}, enabling advanced content generation~\cite{jiangSurveyLargeLanguage2024}, large-scale data analytics~\cite{10.1145/3079628.3079685}, and information-retrieval applications~\cite{razaIndustrialApplicationsLarge2025}. As model scale continues to grow, so do their capabilities, ranging from stronger reasoning and better generalization to improved adaptability across domains. Yet this rapid expansion places increasing pressure on inference infrastructure and compute hardware.

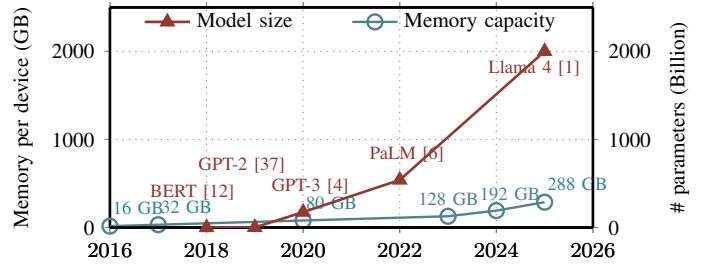
\begin{figure}[t]
\centering
\begin{tikzpicture}
\pgfplotsset{compat=newest}
\begin{axis}[
    width=0.9\columnwidth,
    height=4.5cm,
    xmin=2016, xmax=2026,
    xtick={2016, 2018,2020,2022,2024,2026},
    axis y line*=left,
    ylabel={Memory per device (GB)},
    ymin=0, ymax=2500,
    ytickten={-500,0,500,1000,1500,2000,2500},
    yticklabels={-500,0, 1000,2000},
    ytick style={color=gray},
    y label style={font=\footnotesize},
    tick label style={font=\footnotesize, /pgf/number format/set thousands separator={}},
    line width=1pt,
    every axis plot/.append style={line width=0.9pt},
    legend style={draw=none, fill=none, font=\footnotesize, at={(0.48,1.03)}, anchor=north west},
    grid=both,
    grid style={dotted,gray},
    tick align=outside,
    major tick length=2pt,
    nodes near coords,
    point meta=explicit symbolic,
    every node near coord/.append style={
        font=\scriptsize,
        color=softblue,
        anchor=west,
        xshift=-0.6ex,
        yshift=1.5ex
    },
]
  \addplot+[mark=o,font=\scriptsize, mark size=2.8pt, color={softblue!60!gray}] coordinates {
    (2016,16)  [16 GB]
    (2017,32)  [32 GB]
    (2020,80)  [80 GB]
    (2023,128)
    (2024,192)
    (2025,288) [288 GB]
  };

  \node[font=\scriptsize, color={softblue!60!gray}, anchor=west, xshift=-15pt, yshift=6.4pt]
  at (axis cs:2023,128) {128 GB};
   \node[font=\scriptsize, color={softblue!60!gray}, anchor=west, xshift=-10pt, yshift=6.4pt]
  at (axis cs:2024,192) {192 GB};
  \addlegendentry{Memory capacity}
\end{axis}

\begin{axis}[
    width=0.9\columnwidth,
    height=4.5cm,
    xmin=2016, xmax=2026,
    axis y line*=right,
    ylabel={\# parameters (Billion)},
    ymin=0, ymax=2500,
    ytickten={-500,0,500,1000,1500,2000,2500},
    yticklabels={-500,0, 1000,2000},
    ytick style={color=gray},
    y label style={font=\footnotesize},
    tick label style={font=\footnotesize, /pgf/number format/set thousands separator={}},
    line width=1pt,
    legend style={draw=none, fill=none, font=\footnotesize, at={(0.05,1.03)}, anchor=north west},
    tick align=outside,
    major tick length=2pt,
    nodes near coords,
    point meta=explicit symbolic,
    every node near coord/.append style={
       font=\scriptsize,
        xshift=0.6ex,
        yshift=0.7ex
    },
]
  \addplot+[mark=triangle*,font=\scriptsize, mark size=2.5pt, mark options={fill=softred, draw=softred} , color={softred}] coordinates {
      (2018,0.34)
      (2019,1.5)
      (2020,175)  [GPT-3~\cite{10.5555/3495724.3495883}]
      (2022,540)  [PaLM~\cite{chowdheryPalmScalingLanguage2023}]
      (2025,2000)
  };
  \node[font=\scriptsize, color={softred}, anchor=west, xshift=-25pt, yshift=-6.4pt]
  at (axis cs:2025,2000) {Llama 4~\cite{meta_llama4_behemoth_2025}};
  \node[font=\scriptsize, color={softred}, anchor=west, xshift=-25pt, yshift=13.4pt]
  at (axis cs:2018,0.34) {BERT~\cite{devlinBertPretrainingDeep2019}};
    \node[font=\scriptsize, color={softred}, anchor=west, xshift=-25pt, yshift=23.4pt]
  at (axis cs:2019,1.5) {GPT-2~\cite{radfordLanguageModelsAre2019}};
  \addlegendentry{Model size}
\end{axis}
\end{tikzpicture}
\vspace{-10pt}
\caption{
Model scale increases exponentially while on-device memory growth lags behind.
}
\label{fig:hbm_model_logscale}
\end{figure}

As shown in~\Cref{fig:hbm_model_logscale}, LLM parameter counts have grown exponentially from billions to trillions, pushing weight storage into the terabyte range~\cite{daviesEfficientLlmInference2025}, while GPU memory capacity grows much more slowly. Memory, not compute, has become the primary bottleneck: it determines both the maximum deployable model size and the achievable batch size, and \rv{reducing the weight footprint therefore directly translates into larger batches and higher inference throughput}.

Although lossless compression has proven highly effective in domains such as storage systems and multimedia, its potential for reducing LLM model size and improving inference efficiency remains largely unexplored. \rv{Unlike existing model compression methods such as low rank decomposition~\cite{maLlmprunerStructuralPruning2023,gaoDispllmDimensionindependentStructural2024,yangLacoLargeLanguage2024,huLoraLowrankAdaptation2022,renLowrankPruneandfactorizeLanguage2024,koohpayeganiNolaCompressingLora2023} and quantization~\cite{dettmersQloraEfficientFinetuning2023,heAlphaDecayModulewiseWeight2025,linAwqActivationawareWeight2024,kimOutlierMattersStatistical2025,frantarGptqAccuratePosttraining2022,linAwqActivationawareWeight2024,panSmoothquantAccurateEfficient2023,xiaoSmoothquantAccurateEfficient2023,nagelAdaptiveRoundingPosttraining2020}, lossless compression preserves the exact numerical representation of model weights and therefore guarantees identical inference results. This property is particularly important for the deployment of reasoning-intensive tasks, where lossy techniques can introduce instability and unpredictable accuracy degradation~\cite{wang2024understanding,kalamkarStudyBFLOAT16Deep2019,li2025quantization,liu2025quantization}. Furthermore, lossless compression is orthogonal to existing lossy compression techniques and can be applied on top of well-tuned quantized models to further reduce memory footprint.}

\rv{State-of-the-art lossless compression methods for LLM weights, such as NeuZip~\cite{hao2024neuzip} and DFloat11~\cite{zhang70}, are largely designed for floating-point representations. However, modern LLM deployments increasingly adopt diverse numerical formats, including FP8, INT8, and low-bit or group-wise quantization, which are not well supported by these approaches. As a result, existing techniques do not generalize well across emerging LLM weight formats and leave substantial redundancy unexploited.

This raises two fundamental questions for lossless LLM compression: (1) How much strictly lossless compression is theoretically achievable across both unquantized and modern quantized formats? (2) If significant redundancy exists, how can we design a lightweight, high-throughput GPU execution model that integrates lossless compression directly into inference while preserving bit-exact outputs?
}

To answer the first question, we conduct a comprehensive, model-wide analysis grounded in Shannon’s information theory. Shannon limit provides a fundamental lower bound on the bits required to represent weights drawn from any underlying distribution, independent of numeric format or hardware layout, and therefore quantifies the minimum achievable footprint under any lossless encoding. Our analysis across widely deployed LLMs and numerical formats reveals that, on average, up to 27\% of the stored bits are redundant, indicating significant headroom for footprint reduction without altering weight values (details in~\Cref{sec:entropy_gap_in_existing_llm_compression}). This entropy gap suggests that substantial memory savings, and therefore potentially higher inference throughput, can be unlocked with the right compression and inference integration strategy.

As the entropy analysis reveals substantial redundancy in contemporary LLM weights, we pursue the goal of developing a strictly lossless compression and high-throughput decompression framework for LLM inference. Achieving compression close to the Shannon limit and translating this redundancy into runtime gains, however, is non-trivial: both the codec and its system integration must satisfy stringent constraints imposed by GPU-accelerated LLM execution.

First, \rv{compressed bitstreams produced by conventional entropy codecs are not directly addressable. Most entropy coding schemes generate sequential bitstreams, where decoding must proceed in order and cannot easily jump to arbitrary positions. However, GPU GEMM kernels access weights in a strided, tile-granular pattern rather than sequential order.} Transformer projection layers consume weights in small tiles, often non-contiguous and repeatedly reused across the M-dimension. \rv{This mismatch makes it difficult to directly retrieve the required tiles from a compressed bitstream.} A practical lossless codec must therefore reconstruct tiles on demand, without introducing latency, excessive buffering, or additional global-memory traffic. Only by aligning decompression with the native GEMM tiling structure can the decoding overhead be fully amortized across the compute pipeline.

Second, \rv{a key challenge is that directly applying existing lossless decoders to LLM inference pipelines leads to inefficient execution and poor performance. State-of-the-art compression frameworks~\cite{zhang70,hao2024neuzip} typically follow a decompress-store-compute workflow, where compressed weights are first fully decoded into global memory before being consumed by GEMM kernels. This layer-wise decompression introduces substantial overhead: it requires additional memory capacity to hold the decompressed weights, generates redundant global-memory traffic, and prevents effective overlap between decompression and computation. As a result, naive decompression can become a critical performance bottleneck in real-world LLM serving systems.}

To address these challenges, we begin by analyzing the information content of LLM weights and comparing existing lossless compression families, including dictionary-based methods (LZ77/Zstd), symbol-based entropy codes (Huffman, arithmetic coding), and finite-state entropy coders (rANS/tANS/FSE). %
\emph{Asymmetric Numeral Systems (ANS)} emerges as the only class that simultaneously (1) approaches Shannon-optimal bitrates, (2) supports random access at tile granularity, and (3) enables parallel decoding suitable for GPUs.

Building on this insight, we propose a \emph{tile-level compression with on-the-fly decompression} framework based on ANS that integrates entropy decoding directly into the GPU inference pipeline. \rv{Unlike prior approaches that treat compression as a preprocessing or storage optimization, our design elevates compression to an execution primitive by introducing a compressed-weight execution model in which entropy-coded weight tiles become first-class operands of tensor-core computation. Instead of pre-decompressing weights or introducing layer-level synchronization, compressed tiles are decoded into shared memory precisely when the GEMM microkernel consumes them.} This design enables decompression throughput close to the weight input rate of the baseline GEMM kernel while preserving strict losslessness without affecting model accuracy.

In summary, we make three key contributions:
\begin{itemize}[leftmargin=*]
\setlength\itemsep{-0.5mm}

\item \rv{We perform the first systematic information-theoretic analysis of LLM weights across diverse numerical formats beyond floating-point representations}, revealing a substantial entropy gap to the theoretical Shannon bound. Our results show that, in large models, up to a $\mathbf{5\times}$ reduction in memory footprint is achievable without any accuracy loss.

\item \rv{We propose a compressed-weight execution model for GPU inference that integrates entropy decoding directly into the GEMM pipeline. The design introduces tile-addressable compressed streams, warp-cooperative decoding, and shared-memory reconstruction of weight tiles, enabling decompression and tensor-core computation to proceed concurrently without intermediate global-memory materialization.}

\item \rv{We propose a GPU kernel that integrates seamlessly with standard LLM inference frameworks such as SGLang, enabling larger batch sizes and achieving up to $1.6\times$ higher throughput. Our evaluation further demonstrates that the proposed approach outperforms existing state-of-the-art lossless compression techniques.}

\end{itemize}

\section{Information Redundancy in Contemporary LLM Weight}
\label{sec:entropy_gap_in_existing_llm_compression}

Large language models consist of many transformer layers, each layer performing several projections with high-dimensional matrix multiplications between the weight matrices and the hidden-state activations. These projection weights dominate the static memory footprint of the model and must be frequently accessed from device memory during inference.

\subsection{Heavy-Tailed Distributions in LLM Weights}

LLM weights exhibit heavy-tailed statistical distributions~\cite{nagelAdaptiveRoundingPosttraining2020} and, importantly, the distribution can vary substantially across layers. Most existing studies evaluate compression~\cite{maLlmprunerStructuralPruning2023,gaoDispllmDimensionindependentStructural2024,yangLacoLargeLanguage2024,huLoraLowrankAdaptation2022,renLowrankPruneandfactorizeLanguage2024,koohpayeganiNolaCompressingLora2023} and quantization~\cite{dettmersQloraEfficientFinetuning2023,heAlphaDecayModulewiseWeight2025,linAwqActivationawareWeight2024,kimOutlierMattersStatistical2025,frantarGptqAccuratePosttraining2022,linAwqActivationawareWeight2024,panSmoothquantAccurateEfficient2023,xiaoSmoothquantAccurateEfficient2023,nagelAdaptiveRoundingPosttraining2020} primarily from a numerical-format perspective, emphasizing bit width, scaling, and calibration strategies. However, such approaches do not explicitly capture the underlying data distribution or its cross-layer variability, leaving significant opportunities for further redundancy reduction unaddressed.

To evaluate the potential for further reducing the memory footprint of model weights, we begin with an information-theoretic analysis based on Shannon’s source coding theorem. In information theory, \rv{the Shannon limit defines the theoretical lower bound on the average number of bits required to represent data without loss. This bound is determined by the Shannon entropy of the source distribution, which quantifies the intrinsic information content of the weight values.}

\subsection{Empirical Observations on Model Weights and Numerical Representations}

To compute this bound in practice, we start by computing entropy at the granularity of each individual weight tensor and then aggregating the results across the entire model.
For each layer \(l\), we treat its weight matrix \(W^{(l)}\) as a collection of discrete symbols determined by the chosen numeric format (e.g., FP8 codes, INT4 values, per-channel quantized indices).
We build an empirical histogram over these symbols and estimate the Shannon entropy
\[
\mathcal{H}^{(l)} = -\sum_i p^{(l)}_i \log_2 p^{(l)}_i,
\]
where \(p^{(l)}_i\) is the empirical probability of symbol \(i\) within that tensor.

Because different layers contain different numbers of parameters, we compute the entropy of the full model as a symbol-weighted average of the per-layer entropies:
\[
\mathcal{H}_{\text{model}}
= \frac{\sum_l \mathcal{H}^{(l)} \, |W^{(l)}|}
       {\sum_l |W^{(l)}|},
\]
where \(|W^{(l)}|\) is the number of elements in layer \(l\).
This yields the effective number of bits per weight required under any lossless encoding scheme.

\begin{figure}[t]
    \centering
    \includegraphics[width=\linewidth]{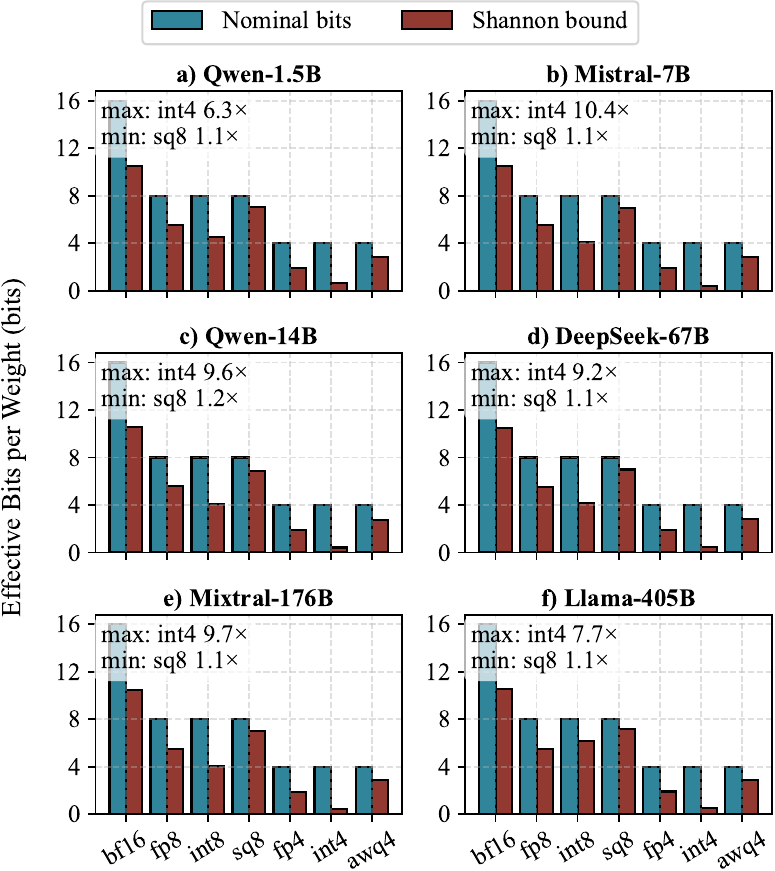}
    \caption{ Remaining entropy gap across models with different data types.}
  \label{fig:entropy_gap}
\end{figure}

We evaluate widely used open-source LLMs ranging from 1.5B to 405B parameters, including Qwen-1.5B~\cite{qwen2.5}, Mistral-7B~\cite{jiang2023mistral7b}, Qwen-14B~\cite{qwen2.5}, DeepSeek-67B~\cite{deepseekai2025deepseekv3technicalreport}, Mixtral-176B~\cite{mistral_mixtral-8x22b_2024}, and Llama-405B~\cite{touvron2023llamaopenefficientfoundation}. Moreover,
we examine the full spectrum of standard numeric formats such as \texttt{bfloat16} (bf16), \texttt{FP8-E5M2} (fp8), \texttt{INT8}, \texttt{FP4-E2M1} (fp4), and \texttt{INT4}, as well as group-quantized formats from \texttt{SmoothQuant}~\cite{xiaoSmoothquantAccurateEfficient2023} (sq8) and \texttt{AWQ}~\cite{linAwqActivationawareWeight2024} (awq4), allowing us to quantify how much redundancy remains in current weight representations and assess how far existing numeric formats are from the theoretical limit.

\Cref{fig:entropy_gap} shows that the effective bits of LLM weights are dramatically lower than their stored bitwidth, often by a factor of two to six. Across all models, \texttt{bfloat16} exhibits about $4\text{--}5$ bits of redundancy, corresponding to a potential $1.5\times$ reduction in size. \rv{Here we refer to \emph{redundancy} as the difference between the stored bitwidth and the estimated entropy of the weight distribution (in bits per weight). For example, if \texttt{bfloat16} stores each weight using 16 bits but the measured entropy of the distribution is approximately 11–12 bits, then about 4–5 bits per weight represent statistical redundancy that can be removed by an optimal lossless coding scheme.}

Even in extremely low-bit representations such as \texttt{INT4} and \texttt{FP4}, substantial redundancy remains because the quantized weight distributions are highly skewed, with only a small subset of symbols appearing frequently due to the heavy-tailed distribution of LLM weights. As shown in \Cref{fig:entropy_gap}, these formats exhibit the largest entropy gaps, with entropy ratios reaching $6\text{--}10\times$, indicating significant unused capacity even at very low bitwidths.

In addition, widely deployed group-quantized formats such as \texttt{SmoothQuant} and \texttt{AWQ} adopt block-based scaling factors to reduce quantization error. While these schemes improve numerical accuracy, they still retain measurable redundancy, typically around $1.1\text{--}1.3\times$ above their entropy bound, due to skewed symbol frequencies and the structured metadata introduced by per-group scaling.

These observations indicate that existing models store far more bits than their intrinsic information content requires, suggesting that {substantial memory savings of up to ten times remain achievable without sacrificing accuracy.}

\begin{table*}[t]
\centering
\caption{Comparison of widely used compression algorithms.
Entropy efficiency measures proximity to the Shannon limit, while streaming capability indicates the smallest granularity at which data can be decoded without global synchronization.}
\label{tab:lossless_compression}
\begin{tabular}{@{}llcc@{}}
\toprule
\textbf{Algorithm / Family} & \textbf{Core Principle} &
\textbf{Entropy Efficiency} &
\textbf{Access Granularity} \\
\midrule
\textbf{gzip (DEFLATE)~\cite{deutschRfc1951DeflateCompressed1996}} & LZ77 + Huffman & 80--90\% & Sequential (per block, $\sim$64\,KB) \\
\textbf{LZ4~\cite{zhangHighRatioCompressionMachineGenerated2023}} & LZ77 (no entropy stage) & $\approx$80\% & Byte-level (continuous) \\
\textbf{Zstandard (Zstd)~\cite{colletRFC8878Zstandard2021}} & LZ77 + FSE (rANS) & 90--95\% & Chunk-level (64\,KB--4\,MB configurable) \\
\textbf{Brotli~\cite{alakuijalaBrotliGeneralpurposeData2018}} & Context + Huffman & 90--95\% & Block-level (windowed) \\
\textbf{Huffman Coding~\cite{moffatHuffmanCoding2019}} & Static symbol code & 90--95\% & Symbol-level (per byte or token) \\
\textbf{Arithmetic Coding~\cite{wittenArithmeticCodingData1987}} & Range interval & 98--99\% & Bit-level (serial) \\
\textbf{BWT / PPM~\cite{effrosPPMPerformanceBWT2002}} & Transform + context & 95--98\% & File-level (global transform) \\
\textbf{rANS / tANS / FSE~\cite{dudaUseAsymmetricNumeral2015}} & Finite-state entropy & \textbf{$>$99\% (near Shannon Limit)} & \textbf{Byte-level (fully streaming)} \\
\bottomrule
\end{tabular}
\label{tab:compression_overview}
\end{table*}

\section{Towards High-Performance LLM Inference with Near-Shannon Limit Compressed Weights}

\label{sec:principle}

\begin{figure}[t]
    \centering
    \includegraphics[width=\linewidth]{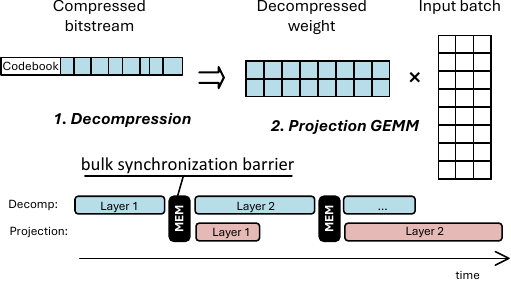}
    \caption{Existing coarse-grain pipelining between decompression and transformer layer computation.
    }
    \label{fig:tile_otfa}
\end{figure}

Incorporating compressed weights into LLM inference naturally consists of two stages: \textbf{(a)~Offline compression}, where weights are compressed during preprocessing using algorithms aggressive enough to approach the Shannon entropy limit; and \textbf{(b)~On-demand decompression at inference}, where compressed weights must be decoded just before the transformer kernels consume them. As shown in~\Cref{fig:tile_otfa}, existing systems such as ZipNN~\cite{hershcovitchZipnnLosslessCompression2025} decompress an entire layer before launching its GEMM. This places decoding on the critical path, introduces a layer-level synchronization barrier, and produces idle compute, redundant global-memory traffic, and memory stalls.

\subsection{Principles for Selecting Lossless Compression and Decompression Algorithms}

Effective lossless compression for LLM inference requires more than a high compression ratio: because GEMM kernels partition projection weights into fine-grained tiles that are repeatedly loaded and reused by the tensor cores, any practical codec must respect this tiled access pattern and integrate seamlessly into the GEMM dataflow rather than treat weights as monolithic tensors.

\noindent\textbf{Principle 1: Near-Shannon bound compression with hardware-realistic throughput}.
LLM inference reads weights at an extremely high bandwidth. A useful lossless codec must therefore compress close to the entropy limit while also decoding fast enough to keep up with GPU memory throughput. Otherwise, decompression becomes the bottleneck, no matter how good the compression ratio.

\noindent\textbf{Principle 2: Tile-granularity access for decompression}.
GEMM kernels consume weights tile by tile, not layer by layer. Most compression algorithms cannot jump to an arbitrary tile without decoding everything before it. This forces full-layer decompression and blocks pipelining. A practical codec must allow each tile to be decoded independently, exactly when the GEMM kernel needs it.

\noindent\textbf{Principle 3: Tight integration of decompression with matrix-multiplication tiling}.
Even with fast, tile-level decoding, decompression must fit directly into the GEMM dataflow. Writing decoded tiles back to global memory wastes bandwidth and breaks overlap. Decompression should instead write directly into shared memory in the same layout used by tensor cores, enabling seamless overlap with computation and avoiding extra memory traffic.

\subsection{Review of Existing Lossless Compression Methods}
Given these three principles, we review classical compression algorithms that achieve excellent ratios in general-purpose data. \Cref{tab:lossless_compression} compares widely used compression schemes, highlighting entropy efficiency and access granularity.

\subsubsection{Dictionary-based compressors (gzip, LZ4, Zstd).}
LZ77-style schemes rely on sequential pointer chasing through a sliding dictionary, which prevents random-access decoding of a single weight tile without rebuilding all prior state. \rv{Their compression efficiency also degrades at tile granularity, since effectiveness relies on large dictionary contexts.} Both properties violate the tile-granularity constraint for pipelined GEMM execution.

\medskip

\subsubsection{Symbol-based codecs (Huffman, arithmetic coding).}
Huffman coding is fast but limited by integer-length codes, leaving nontrivial gaps to the Shannon limit. Arithmetic coding achieves near-optimal entropy efficiency, but its bit-serial state machine prevents parallel decoding and restricts throughput to only a few GB/s. Both methods fail to meet the two requirements: they neither support tile-level random access nor scale to HBM-level bandwidth during inference.

\medskip

\subsubsection{Finite-state entropy coders (rANS, tANS, FSE).}
Finite-state entropy coding retains the near-Shannon efficiency of arithmetic coding but replaces serial interval updates with table-driven state transitions, enabling byte-level streaming and parallel decoding at tens to hundreds of GB/s while keeping each tile independently decodable. It is the only codec class that meets both constraints required for high-throughput LLM inference.
{Prior work, such as DietGPU~\cite{dietgpu2025}, provides competitive warp-cooperative rANS decoders, offering a strong foundation with the potential for further performance optimization and tight integration with the GEMM execution pipeline.}

\section{On-the-fly Decompression}
\label{sec:ans_on_the_fly}
\Cref{fig:tile_otfb} illustrates our on-the-fly decompression execution model, which replaces coarse, layer-level decoding with a continuous, tile-aligned dataflow.

The compressed weight bitstream is partitioned into tile-sized ANS substreams, with each tile’s starting offset recorded in a lightweight index table.
All tiles within the same layer share a compact ANS codebook, which is constructed from the layer’s aggregated weight distribution.
Sharing a layer-wide codebook maximizes statistical coverage of the underlying distribution while keeping metadata overhead minimal. Each tile can then be decoded independently, exactly at the moment the GEMM kernel needs it, without scanning preceding tiles. The runtime workflow consists of three tightly coupled stages:

\begin{figure}[t]
    \centering
    \includegraphics[width=\linewidth]{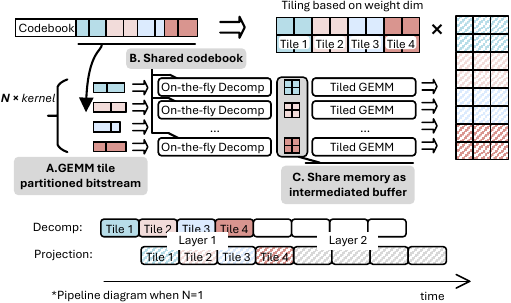}
    \caption{Tile-aligned on-the-fly decompression, partitions weights into fine-grained tiles, decodes them on demand, and overlaps decompression with GEMM execution using shared memory buffers.
    }
    \label{fig:tile_otfb}
\end{figure}

\noindent\textbf{A. Tile-partitioned bitstreams.}
During offline preprocessing, each weight matrix is partitioned into tile-aligned substreams matching the GEMM tile size executed on the SMs. Each tile is entropy-encoded as an independent ANS bitstream using a shared per-layer codebook, and a compact offset entry is stored in the tile index table to enable direct tile access.

\noindent\textbf{B. On-the-fly ANS decompression.}
At runtime, multiple ANS decoder kernels execute on the SMs, reconstructing weight tiles directly into their on-chip shared memory. This avoids writing decompressed weights back to global memory, substantially reducing global memory bandwidth consumption.

\noindent\textbf{C. Coupled GEMM execution.}
As soon as a tile is decoded in shared memory, it is immediately consumed by the GEMM kernel in its required swizzled layout.
A double-buffered shared-memory workspace ensures that, while one tile is being used for computation, the next tile is being decoded in Tile-aligned on-the-fly .

By aligning decompression with GEMM tile consumption, the proposed design removes layer-level synchronization barriers and eliminates intermediate global-memory traffic. As shown in the timeline comparison, computation for layer~$i$ can begin as soon as the first tiles are ready, while later tiles are decoded concurrently, achieving sustained line-rate throughput with negligible overhead.

\begin{figure*}[t]
    \centering
    \includegraphics[width=0.95\linewidth]{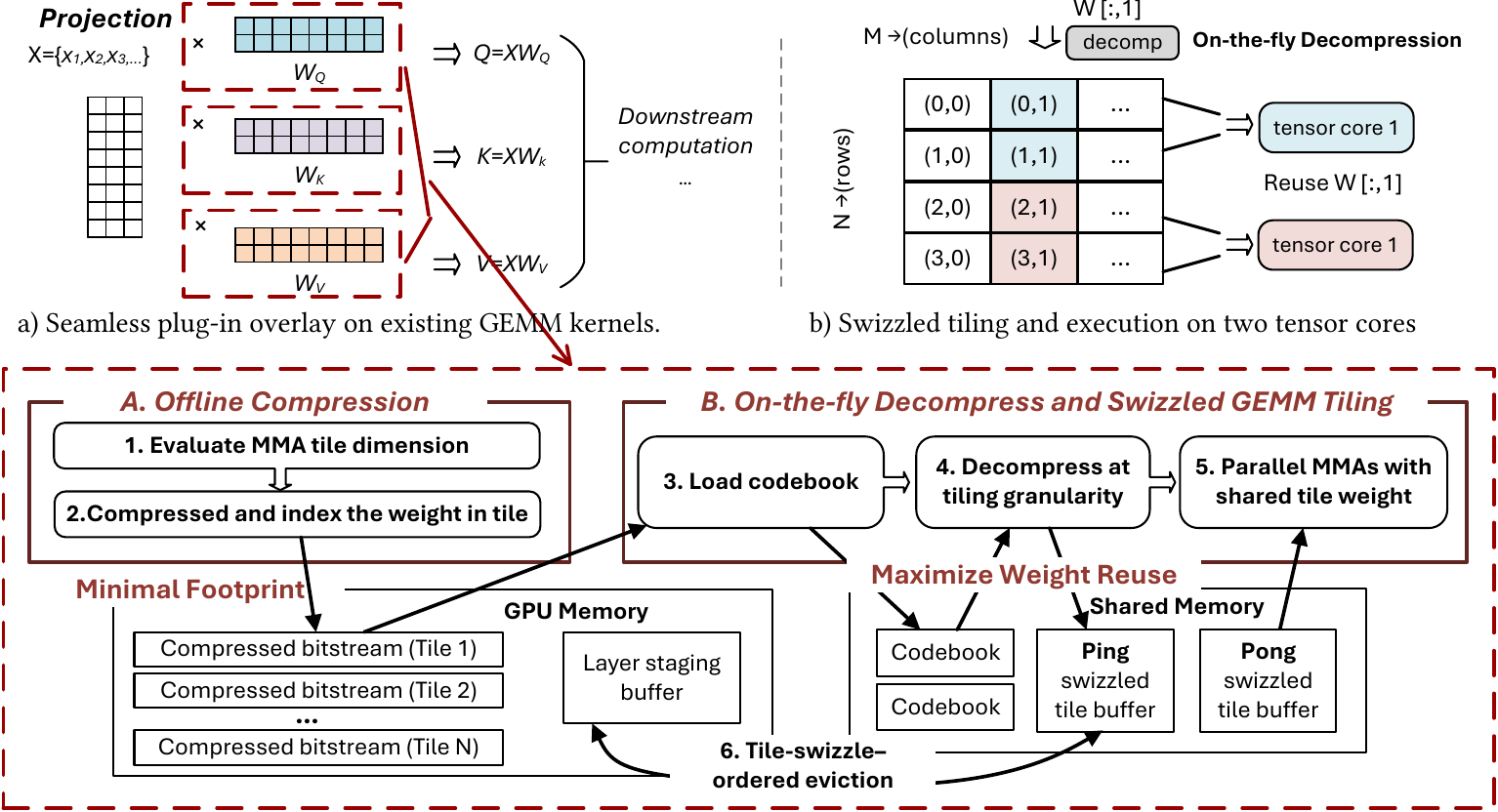}
    \caption{
    GPU kernel for on-the-fly decompression and swizzled GEMM execution.
    (a) Projection layers seamlessly integrate the decompression kernel.
    (b) Offline compression minimizes footprint, while on-the-fly decompression and shared-memory swizzling maximize weight reuse and pipeline overlap.
    }
    \label{fig:gpu_kernel_final}
\vspace{-8pt}
\end{figure*}

\section{GPU Implementation and Optimizations}
\label{sec:gpu_kernel}

In this section, we describe how to integrate the on-the-fly decompression mechanism into GPU execution pipelines.

\subsection{Design Overview}

Figure~\ref{fig:gpu_kernel_final} illustrates the complete workflow of the proposed GPU runtime kernel.
Building on state-of-the-art GEMM libraries such as CUTLASS, we develop a lightweight plugin-style kernel that can override existing projection operators with minimal changes to library-level GEMM implementations and provides a wrapper for direct invocation in PyTorch.

The design consists of two complementary stages: the offline compression stage, minimizing the static device-memory footprint, and the on-the-fly decompression and swizzled GEMM stage,  maximizing runtime efficiency during inference.
For the decompression backend, we build upon the ANS kernel implementation from the open-source DietGPU library~\cite{dietgpu2025}, adapting and extending it to support tile-granular decoding and integration with modern tensor-core GEMM pipelines.

\subsection{Tile Addressable ANS Offline Compression}

In the preprocessing phase, each projection matrix ($W_Q$, $W_K$, $W_V$) is first profiled to determine the optimal tensor-core tiling geometry (e.g., $128{\times}32$, $256{\times}64$, $128{\times}128$ and etc.) for downstream GEMM execution.
We then aggregate the weight statistics across the entire layer to construct a shared, compact ANS codebook that captures the dominant distributional structure while amortizing codebook overhead of independent compression on chunks.
After the codebook is established, the weight tensor is partitioned into tiles \rv{aligned with the GEMM tiling geometry}. Each tile is then entropy-encoded using ANS with an independently initialized state while sharing the same per-layer codebook.
Because ANS supports arbitrary initial states without losing compression efficiency, each tile becomes a fully self-contained substream that can be decoded independently at inference time.

The resulting compressed bitstreams and their tile-offset metadata are stored in GPU memory, producing a compact representation while preserving the exact tile boundaries required by GEMM kernels.
Importantly, this stage is \emph{entirely offline} and independent of framework execution. The compressed model can be loaded as a direct drop-in replacement for standard weights, enabling our system to act as a lightweight plugin atop existing LLM inference frameworks such as PyTorch or custom CUDA runtimes.

\begin{algorithm}[t]
\SetAlgoVlined
\DontPrintSemicolon
\SetKwProg{Fn}{Function}{}{}
\SetKwFunction{DecodeTile}{RansDecodeTile}
\SetKwFunction{GemmTile}{GemmTile}

\caption{Fused rANS Decompression and GEMM Tile Computation \label{alg:fused_rans_gemm}}

\KwIn{Compressed bitstreams $\mathcal{B}$, global rANS decode table $T$, matrices $B$ and $C$}
\KwOut{Updated output tile $C$}

\BlankLine
\textbf{Shared memory:}\\
\Indp
\quad Decode table $\tilde{T}$ (copied from $T$)\;
\quad Double-buffered decompressed tiles $A^{(0)}, A^{(1)} \in \mathbb{R}^{M \times K}$\;
\quad Tile $B^{(k)} \in \mathbb{R}^{K \times N}$\;
\quad Atomic Flags $\mathsf{ready}[2]$ indicating buffer readiness\;
\Indm

\BlankLine
Copy global decode table $T$ into shared memory $\tilde{T}$\;
Initialize $\mathsf{ready}[0] \leftarrow 0$, $\mathsf{ready}[1] \leftarrow 0$\;

\BlankLine
\For{$k = 0$ \KwTo $K_{\mathrm{tiles}}-1$}{
    $b \leftarrow k \bmod 2$   \tcp*{current buffer index}
    $p \leftarrow 1-b$         \tcp*{previous buffer}

    \BlankLine
    \textbf{Warp 0: rANS decompression into $A^{(b)}$}\;
    \Indp
    \DecodeTile{$A^{(b)},\, \mathcal{B}[k],\, \tilde{T}$}\;
    $\mathsf{ready}[b] \leftarrow 1$\;
    \Indm

    \BlankLine
    \textbf{Warps 1..W: GEMM tile compute}\;
    \Indp
    Load $B^{(k)}$ from global memory\;

    \If{$k = 0$}{
        Wait until $\mathsf{ready}[b] = 1$  \tcp*{first tile}
        \GemmTile{$A^{(b)}, B^{(k)}, C$}\;
        $\mathsf{ready}[b] \leftarrow 0$\;
    }
    \Else{
        Wait until $\mathsf{ready}[p] = 1$  \tcp*{consume previous buffer}
        \GemmTile{$A^{(p)}, B^{(k)}, C$}\;
        $\mathsf{ready}[p] \leftarrow 0$\;
    }
    \Indm

    \BlankLine
    \textbf{Block-wide synchronize}\;
}

\BlankLine
\Return{$C$}

\BlankLine

\end{algorithm}

\begin{algorithm}[t]
\SetAlgoVlined
\DontPrintSemicolon
\SetKwProg{Fn}{Function}{}{}
\SetKwFunction{DecodeTile}{RansDecodeTile}
\SetKwFunction{GemmTile}{GemmTile}

\caption{Warp-cooperative rANS tile decompression\label{alg:rans_decode_tile}}

\Fn{\DecodeTile{$A,\, \text{stream},\, \tilde{T}$}}{
    Initialize rANS state $s$ for each lane\;
    \For{$i = 0$ \KwTo $S_{\mathrm{lane}}-1$}{
        $x \leftarrow s.\mathrm{value} \bmod R$ \tcp*{low bits}
        $\sigma,\; f,\; c \leftarrow \tilde{T}[x]$  \tcp*{shared-memory lookup}
        $w \leftarrow \mathrm{DecodeSymbol}(\sigma)$\;
        Compute $(r,c)$ for symbol index and write $A[r,c] \leftarrow w$ \tcp*{$A$ is a shared-memory tile}

        \BlankLine
        $s.\mathrm{value} \leftarrow f \cdot \lfloor s.\mathrm{value}/R \rfloor + (x - c)$\;
        \While{$s.\mathrm{value} < \mathrm{renorm\_thresh}$}{
            $u \leftarrow$ Load 32-bit chunk (coalesced)\;
            $s.\mathrm{value} \leftarrow (s.\mathrm{value} \ll 32) \mid u$\;
        }
    }
}
\end{algorithm}

\subsection{On-the-Fly Decompression Pipelined with GEMM}

After the offline compression stage, the compressed weight tensors and their metadata are loaded into GPU memory and used directly during inference. During execution, decompression is performed at the same tile granularity as the GEMM computation and executed in a fused manner for efficient overlap.

\subsubsection{Fused Tile Aligned Kernel Design}
\Cref{alg:fused_rans_gemm} details the fused execution model that underlies our high-performance ANS decompression path. The kernel consists of two cooperating components: a warp-cooperative rANS decoding kernel and a tile-level GEMM microkernel.

At the beginning of each layer, the rANS decode table is loaded into shared memory by the decoding warp on each SM, enabling low-latency, high-bandwidth table lookups during decompression.

During execution, warp~0 fetches compressed weight tiles from global memory and decodes them in a streaming manner. The decoded symbols are written \emph{directly} into a shared-memory tile $A^{(b)}$, with indices generated in lockstep with the interleaved decode order. This avoids materializing decompressed weights in global memory and significantly reduces global-memory traffic. The rANS decoder therefore acts as a producer that reconstructs weight tiles directly into shared memory, while the remaining warps in the thread block immediately consume the tile for tensor-core GEMM computation.

A double-buffered pipeline overlaps decompression of tile $k{+}1$ with GEMM computation on tile $k$. \rv{Producer and consumer warps synchronize through shared-memory atomic state flags implemented with \texttt{cuda::atomic\_ref<int, cuda::thread\_scope\_block>}, ensuring correct ordering with minimal performance overhead.} The detailed scheduling of producer and consumer warps on disjoint hardware pipes, and the regime in which decode is fully hidden, are analyzed in \Cref{subsec:overlap_regime}.

\subsubsection{Tile Swizzle and Deterministic Eviction}
For large-dimension GEMM, the kernel follows a fixed tile-swizzle traversal order to maximize data reuse and balance shared-memory pressure across thread blocks. Because this access sequence is deterministic and determined by the GEMM tiling schedule, the working set of active tiles may temporarily exceed the available shared-memory capacity.
In such cases, \rv{decoded tiles are evicted following the deterministic order induced by the tile swizzle traversal. The notion of “recency” is therefore defined by the swizzle access order rather than runtime reuse tracking, allowing the eviction sequence to be determined statically without additional bookkeeping.} When a tile is evicted from shared memory, its decompressed form is temporarily stored in a small decompression buffer to avoid redundant decompression if the tile is accessed again shortly thereafter, while the compressed representation remains in global memory.

\subsubsection{Parallel Warp-Cooperative rANS Decoding}

\Cref{alg:rans_decode_tile} presents the warp-cooperative rANS decoding algorithm used to reconstruct compressed weight tiles on the GPU. Because the rANS state machine is inherently sequential, \rv{we expose parallelism by partitioning each tile’s compressed bitstream into $R$ independent substreams. Each warp lane maintains its own rANS state and processes one substream, allowing the serial decoding process to be distributed across the warp.}

For each decoding step, the lane extracts the low bits of the current rANS state to determine the next symbol, performs a shared-memory lookup to retrieve the corresponding table entry, and writes the decoded value directly into the shared-memory.

\rv{To maintain correctness of the rANS automaton, each lane independently updates and renormalizes its state by reading additional bits from the compressed stream when the state falls below the renormalization threshold. Because the compressed streams are interleaved across lanes, these renormalization loads are naturally coalesced in global memory, preserving high memory throughput.
This warp-level interleaving enables a single SM to decode multiple rANS streams concurrently while maintaining the correctness of the rANS state transitions. As a result, the decoder achieves high parallel efficiency while reconstructing tiles directly into shared memory for immediate consumption by the GEMM kernel.
}

\subsubsection{Tile-Level GEMM Microkernel}

\Cref{alg:gemm_tile} then illustrates the complementary consumer stage. Here, the GEMM microkernel reads the decompressed weight tile $A$ directly from shared memory and multiplies it with a $K\times N$ activation tile $B$. Because the weight tile is already resident in shared memory accessible to the warp, all accesses to $A[m,k]$ are single-cycle, eliminating the bandwidth demands and cache thrashing associated with repeatedly loading large weight matrices from global memory.

\rv{We build upon CUTLASS to obtain the flexibility required to support the diverse numeric formats and widely used quantization schemes. Moreover, CUTLASS exposes programmable tensor-core tiling, warp scheduling, and memory layouts, allowing us to integrate tile-level ANS decompression directly into the GEMM pipeline, but our proposed design is not tied to CUTLASS itself.}

A key consequence of this organization is that decode and compute share the same on-chip shared-memory footprint, so each weight is decoded exactly once and never reloaded. Combined with the producer-consumer scheduling analyzed in \Cref{subsec:overlap_regime}, the kernel sustains GEMM at full tensor-core speed while keeping the decoder on the critical path only when batch size is small.

\subsection{Pipeline Overlap and the Batch-Size Regime}
\label{subsec:overlap_regime}

The producer (rANS decode) and consumer (tensor-core matrix multiply) execute on physically disjoint pipelines within each streaming multiprocessor: decode warps exercise only the integer and load/store units (probability-table loads, rANS state updates, shared-memory stores into the operand slab), while matrix-multiply warps issue only shared-memory matrix loads and tensor-core multiply–accumulate instructions on the separate tensor pipeline. The warp scheduler co-issues them in the same cycle, and a four-stage shared-memory ring buffer lets decode run several sub-tiles ahead, so each consumer step finds its operands already resident.

The effectiveness of this overlap scales with batch size. Decode cost per sub-tile is approximately constant, dominated by probability-table accesses and renormalization reads; matrix-multiply cost per sub-tile grows with the $M$-rows processed by each thread block, since one decoded $B$ fragment is reused across all $M$-blocks. At small batch sizes the tensor pipeline drains faster than the decoder can produce the next sub-tile and the kernel is decoder-bound; at large batch sizes matrix-multiply work dominates, the consumer never stalls, and tensor utilization recovers toward the uncompressed baseline. The fused kernel therefore achieves near-complete overlap in the high-batch regime and degrades gracefully to an unfused decode-then-GEMM schedule at low batch.

\begin{algorithm}[t]
\SetAlgoVlined
\DontPrintSemicolon
\SetKwProg{Fn}{Function}{}{}
\SetKwFunction{DecodeTile}{RansDecodeTile}
\SetKwFunction{GemmTile}{GemmTile}
\caption{Shared-memory GEMM microkernel\label{alg:gemm_tile}}
\Fn{\GemmTile{$A, B, C$}}{
    Each warp reads its fragment of the shared-memory weight tile $A$ and accumulates into $C$\;
    \For{$m,n,k$ in tile loops}{
        $C[m,n] \mathrel{+}= A[m,k] \cdot B[k,n]$ \tcp*{weights $A[m,k]$ come from shared memory}
    }
}
\end{algorithm}

\subsection{Global Memory Access Reduction}

Consider a tiled GEMM $C[M \times N] = A[M \times K] \cdot W[K \times N]$ with blocking
$(M_t, N_t, K_t)$ and a compute kernel that stages one $W$–tile of size $K_t \times N_t$
into shared memory while iterating over the $M$ dimension.

Without loss of generality, we assume the projection matrix $W \in \mathbb{R}^{K \times N}$ does not fit in cache at all, so every time the kernel advances along the $M$-dimension it has to reload the relevant tiles of $W$ from HBM. Since there are $M / M_t$ tiles along the M-dimension,  each element of $W$ is therefore loaded once per $M_t$-tile, so the global-memory traffic for $W$ becomes
\begin{align}
V_B^{\text{uncompressed}} = \frac{M}{M_t} \cdot K N.
\end{align}
In contrast, with our on-the-fly ANS kernel, each element of $W$ is fetched once in compressed form and then reused from shared memory.
\begin{align}
\label{eq:gpu_kernel}
V_B^{\text{on-the-fly decomp}} = (\frac{M}{M_t} + \alpha - 1) \cdot K N .
\end{align}

Despite the fact that decompression throughput may not fully match the raw memory bandwidth of GPUs, our design can still sustain high GEMM throughput. By decoding tiles directly into shared memory and overlapping decompression with computation through carefully orchestrated pipeline scheduling, the system eliminates the repeated global-memory loads that dominate baseline execution. Moreover, the on-the-fly decompression stage introduces no additional global-memory pressure, ensuring that reductions in memory traffic translate directly into end-to-end performance gains.

\section{Evaluation}
\label{sec:evaluation}

We evaluate the proposed \emph{on-the-fly decompression} design along three objectives:
\begin{enumerate}[leftmargin=*, topsep=2pt, itemsep=-1ex, partopsep=2ex, parsep=1ex]

\item \textbf{Entropy characterization and compression efficiency.}
        Quantify the entropy of weights across representative LLMs, evaluate achievable compression ratios, and measure the gap between practical rANS-based coding and the Shannon bound.

\item \textbf{GPU system-level performance.}
        Benchmark end-to-end inference performance with on-the-fly decompression under realistic GPU memory budgets, batch sizes, and sequence lengths.
\item \textbf{Microbenchmark of the proposed technique.}
      Evaluate the efficiency and sensitivity of our design across different GEMM dimensions and under both prefill and decode settings.
\end{enumerate}

\subsection{Experimental Setup}

\noindent\textbf{Evaluated LLM models.}
To evaluate the effectiveness and generality of our approach, we conduct experiments across a diverse set of widely used open-source large language models spanning different scales and architectures. The evaluated models include \texttt{Qwen-1.5B} (Qwen2-1.5B), \texttt{Mistral-7B} (Mistral-7B-v0.3), and \texttt{Qwen-14B} (Qwen-14B), which represent commonly deployed dense transformer configurations. To further examine scalability to larger models, we include \texttt{DeepSeek-67B} (DeepSeek-LLM-67B) and \texttt{Llama-3.1-405B}. In addition, we evaluate mixture-of-experts architectures using \texttt{Mixtral-8x22B} (Mixtral-176B total parameters). These models collectively span parameter scales from 1.5B to 405B and cover both dense transformer architectures and modern MoE designs.

\noindent\textbf{Numeric formats.}
Across these models, we evaluate a broad spectrum of widely adopted numeric formats, including \texttt{bfloat16} (bf16), \texttt{FP8-E5M2} (fp8), \texttt{INT8}, \texttt{FP4-E2M1} (fp4), and \texttt{INT4}, as well as state-of-the-art group-quantized formats used in \texttt{SmoothQuant}\cite{xiaoSmoothquantAccurateEfficient2023} (sq8) and \texttt{AWQ}\cite{linAwqActivationawareWeight2024} (awq4). This range of representations allows us to evaluate the compressibility and runtime performance of our method across both standard floating-point formats and modern quantized weight representations used in LLM inference.

\noindent\textbf{Hardware platforms}.
All compression-size and runtime performance experiments for our GPU kernels are conducted on two GPU servers to demonstrate the portability of our design across different GPU generations. The first server is equipped with eight NVIDIA A100 GPUs (80 GB HBM2e, 2 TB/s peak bandwidth). \rv{We further evaluate the performance on NVIDIA Hopper H200 GPUs.} Experiments are implemented using PyTorch 2.5.1 and CUDA 12.1, with the CUTLASS-based GEMM baselines across platforms to ensure fair comparison.

\begin{table*}[t]
  \centering
  \begin{threeparttable}
  \begin{tabularx}{1\linewidth}{lYYYYYYYY}
    \toprule
    Model & \makecell{Sequence\\Length}& Variant &  Weight Mem (GB) & KV Mem (GB) & Total Mem (GB) & Max Batch & Throughput (Token/s) &  Median TPOT (ms)  \\
\midrule
    \multirow{4}{*}{\makecell{Qwen-14B \\ Budget: 80 GB\\(Single NVIDIA A100 GPU)}}
      & \multirow{2}{*}{1024}
      & Uncompressed                  & 27.5 & 44.1  & 75  & 47 & \textcolor{black}{1131} & \textcolor{black}{71} \\
      &                     & Ours    & 18.1 & 56.3  & 75  & \textbf{60 (\bm{$1.3\times$})} & \textcolor{black}{1217} & \textcolor{black}{81} \\
\\[-7pt]
\cline{2-9}
\\[-7pt]
      & \multirow{2}{*}{2048}
      & Uncompressed                  & 27.5  & 43.1  & 74  & 23 & \textcolor{black}{548} & \textcolor{black}{112}\\
      &                     & Ours    & 18.1  & 56.3  & 75  & \textbf{30 (\bm{$1.3\times$})} & \textcolor{black}{651}  & \textcolor{black}{125}\\
\midrule
    \multirow{4}{*}{\makecell{ Mixtral-176B \\ Budget: 320 GB\\ (Four NVIDIA A100 GPU)}}
      & \multirow{2}{*}{1024}
      & Uncompressed                  & 261.9  & 26.3  & 304  & 20 & \textcolor{black}{241} & \textcolor{black}{110}\\
      &                     & Ours    & 163.7  & 124.6  & 304  & \textbf{95 (\bm{$4.8\times$})} & \textcolor{black}{391} & \textcolor{black}{170} \\
\\[-7pt]
\cline{2-9}
\\[-7pt]
      & \multirow{2}{*}{2048}
      & Uncompressed                  & 261.9  & 26.3  & 304  & 10 & \textcolor{black}{190} & \textcolor{black}{213} \\
      &                     & Ours    & 163.7  & 123.4  & 304  & \textbf{47 (\bm{$4.7\times$})} & \textcolor{black}{257}  & \textcolor{black}{318} \\
    \bottomrule
  \end{tabularx}
  \end{threeparttable}
  \caption{Illustration of memory footprint before and after lossless compression.
  Compression reduces the weight of memory,  freeing capacity for larger KV-cache and enabling larger effective batch sizes and throughput at the same sequence length.}
  \label{tab:compression_memory_throughput}
\end{table*}

\begin{figure}[t]
    \centering
    \includegraphics[width=\linewidth]{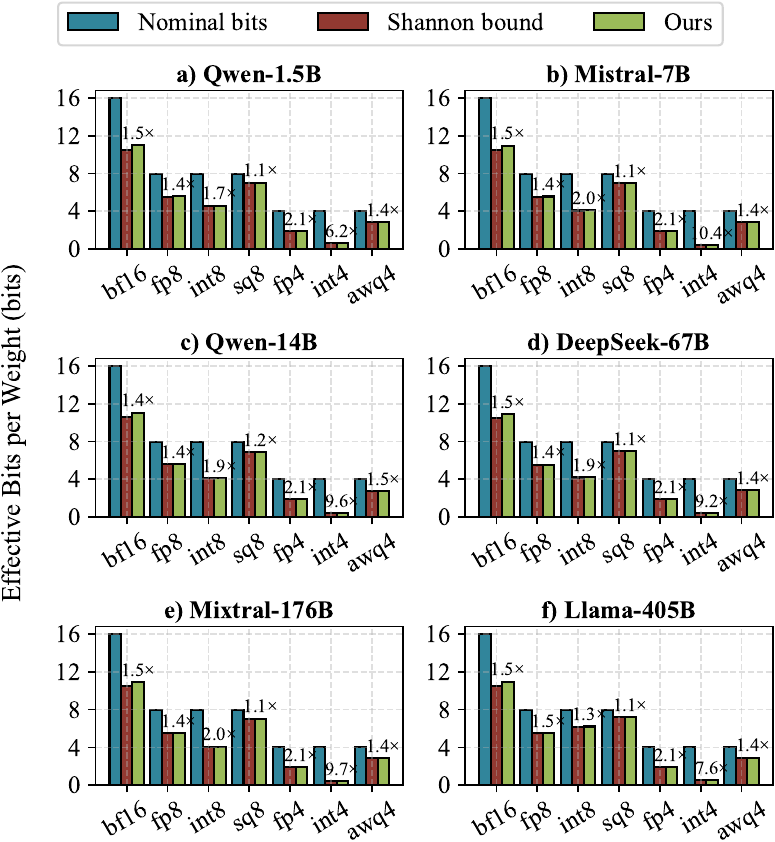}
    \caption{Effective bit rates of tile-level ANS compression relative to Shannon entropy bounds.}
  \label{fig:entropy_comprare}
\end{figure}

\subsection{Entropy Characterization and Compression Bound}

\Cref{fig:entropy_comprare} provides an overview of the effective bits-per-weight achieved by our tile-level ANS compression compared against both the nominal storage formats and the Shannon entropy bound across the evaluated models. We have the following observations.

First, across all models and datatypes, including bf16, fp8, int8, group-quantized formats (sq8, awq4), and low-bit representations (fp4, int4), our ANS bitrates closely track the Shannon bound, typically within 0.01–0.05 bits. This near-perfect overlap indicates that the ANS encoder captures essentially all statistical redundancy in the weight distribution. The agreement is particularly tight for lower-bit formats such as int8, sq8, fp4, and int4, where the symbol alphabet is small and the empirical distribution is sharply peaked, allowing ANS to encode nearly at the theoretical optimum.

Second, the only consistent deviation appears for bf16, where the 16-bit symbol alphabet leads to a broader histogram and a larger frequency table. The finite-precision normalization required for 212-entry frequency tables introduces marginal overhead (roughly 0.1–0.2 bits), which is expected for high-cardinality distributions and well within the limits predicted by finite-precision ANS theory. Importantly, even in this worst case, our ANS bitrate remains within 1.1–1.5$\times$ of nominal precision, significantly closer to the entropy limit than any existing lossless scheme.

Third, the figure highlights that nominal storage formats substantially over-allocate bits relative to the intrinsic information content of the weights. For example, bf16 weights often exhibit an effective entropy of only 10–12 bits, int8 typically compresses to 4–5 bits, and even aggressively quantized formats such as int4 retain only 0.6–1.0 bits of true entropy. These gaps are consistent across all model scales, from Qwen-1.5B to Llama-405B, demonstrating that the heavy-tailed, highly structured nature of transformer weight distributions persists uniformly across architectures and sizes.

\begin{figure*}[t]
    \centering
\begin{minipage}[t]{0.49\linewidth}

    \includegraphics[width=\linewidth]{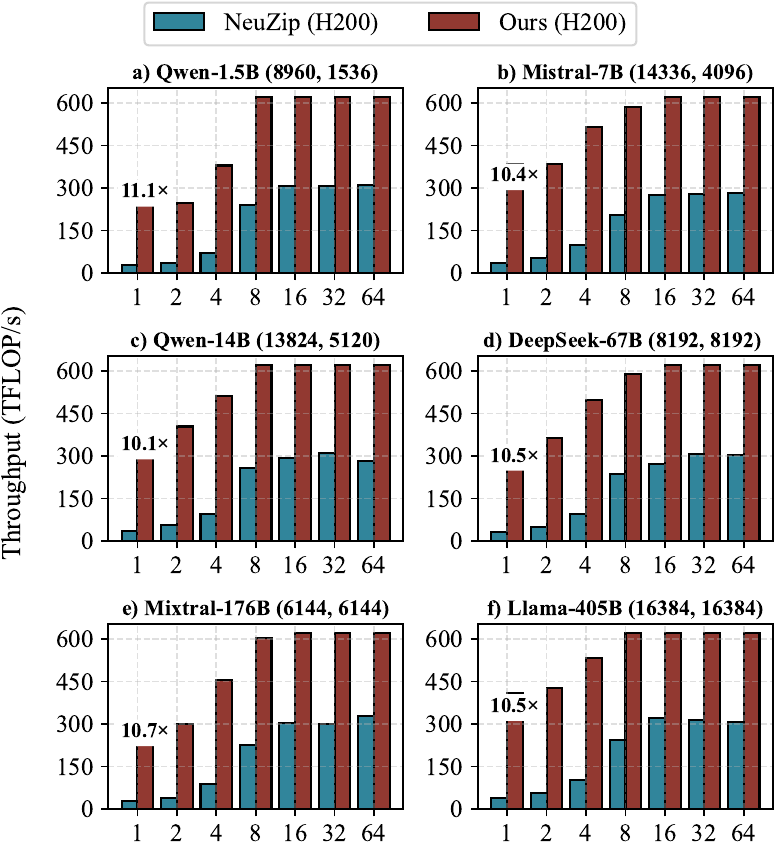}
    \caption{Comparison between our on-the-fly decompression path and the NeuZip~\cite{hao2024neuzip} baseline across varying batch sizes on NVIDIA H200. Weight matrix dimensions are given as $(a,b)$}
    \label{fig:neuzip_cutlass}
\end{minipage}
\hfill
\begin{minipage}[t]{0.49\linewidth}
    \centering
    \includegraphics[width=\linewidth]{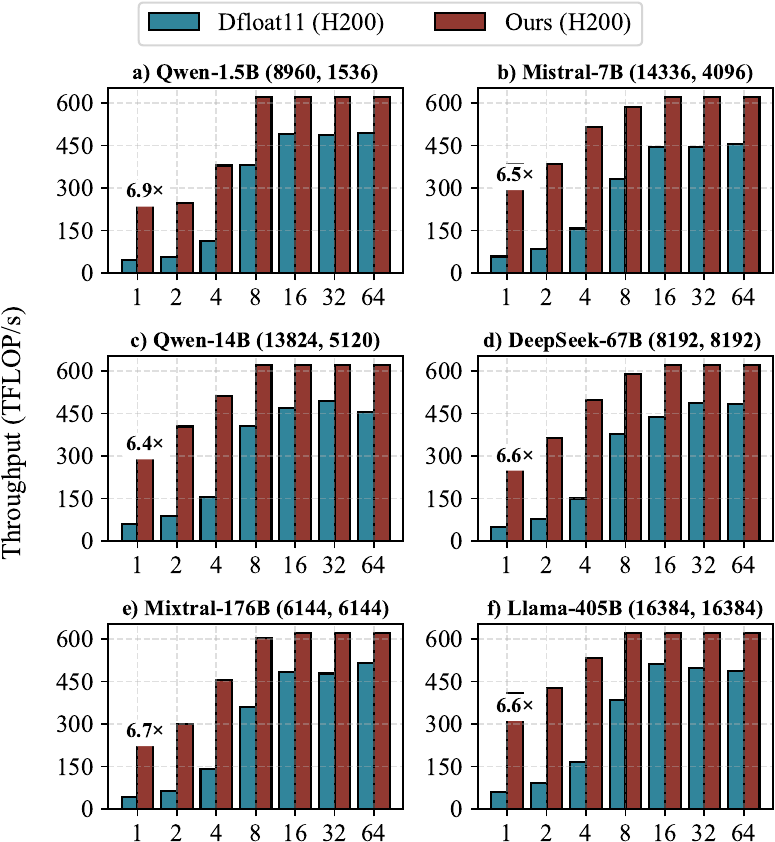}
    \caption{Comparison between our on-the-fly decompression path and the DFloat11~\cite{zhang70} baseline across varying batch sizes on NVIDIA H200. Weight matrix dimensions are given as $(a,b)$}
      \label{fig:dfloat11_cutlass}
\end{minipage}
\end{figure*}

\subsection{End-to-End Performance Gains from Reduced Weight Memory Footprint}

We evaluate the end-to-end inference impact of our reduced memory footprint under realistic device-memory budget constraints using the SGLang serving framework. \rv{We modify SGLang~\cite{zheng2024sglang} to use our ANS-enabled GEMM backend for dense matrix multiplications, comparing it to the default CUTLASS-based kernels in the SGLang runtime.}

Because our approach significantly reduces the memory footprint of model weights, it allows substantially larger query batches to fit within a fixed GPU memory budget. We therefore evaluate the resulting end-to-end throughput improvements under realistic inference workloads. Table~\ref{tab:compression_memory_throughput} reports the memory breakdown, maximum feasible batch size, achieved throughput, and median time-per-output-token (TPOT) for Qwen-14B and Mixtral-176B across two representative sequence lengths (1024 and 2048). \rv{For Mixtral-176B, multi-GPU inference is implemented using expert parallelism (EP) across four GPUs. The reported throughput corresponds to measured execution time under SGLang’s batching scheduler running on GPUs.}

\rv{Although our on-the-fly decompression kernel introduces additional computation compared to a pure CUTLASS GEMM kernel used in existing LLM serving systems, reducing the weight footprint directly increases the effective batch size and enables higher serving throughput. For example, on a single A100 with sequence length 1024, Qwen-14B increases the maximum batch size from 47 to 75, improving throughput from 1131 to 1217 tokens/s (1.1$\times$). For longer sequences (2048), throughput increases from 548 to 651 tokens/s (1.2$\times$).

The effect is more pronounced for larger models. On four A100 GPUs, Mixtral-176B increases the feasible batch size from 20 to 95 at sequence length 1024, resulting in a throughput improvement from 241 to 391 tokens/s (1.6$\times$). At length 2048, throughput improves from 190 to 257 tokens/s (1.4$\times$). These results highlight an important system-level effect: compression fundamentally shifts the bottleneck of LLM inference from device memory capacity to compute throughput. The gains come from the reduced weight memory, which allows more requests to share the KV-cache capacity, improving batching efficiency under realistic workloads.

We also report the median TPOT to capture the latency impact of the compressed execution path. {Our design primarily targets throughput-oriented serving workloads by increasing the effective batch capacity under a fixed memory budget}. While TPOT slightly increases due to the additional on-the-fly decompression cost, the increase is modest compared to the substantial throughput gains enabled by larger batching for the throughput-oriented LLM model serving.
}

\subsection{Comparison to SOTA Lossless Compression LLM}

\rv{
We compare our fused on-the-fly decompression pipeline with two recent lossless LLM compression systems, NeuZip~\cite{hao2024neuzip} and DFloat11~\cite{zhang70} on NVIDIA H200 hardware. We evaluate six representative LLMs of different scales, and each subplot reports the achieved throughput in TFLOP/s as the batch size increases from 1 to 64 of a single layer of projection with 4096 input tokens. The annotation in each plot indicates the peak relative improvement of our method compared with baseline.

Both prior approaches target floating-point formats (e.g., FP16/BF16) and perform layerwise decompression, where an entire compressed layer must first be reconstructed in global memory before GEMM execution. This introduces additional memory traffic and synchronization overhead.
In contrast, our method decodes compressed weights at tile granularity and feeds them directly into the GEMM pipeline. This eliminates global-memory materialization of decompressed layers and overlaps decompression with tensor-core computation.

\Cref{fig:neuzip_cutlass} and \Cref{fig:dfloat11_cutlass} show that our approach outperforms both baselines across all evaluated models and batch sizes on NVIDIA H200. Compared with NeuZip, the fused pipeline achieves up to $\sim10\times$ higher throughput, while outperforming DFloat11 by $\sim6$--$7\times$. These gains show that tight integration of entropy decoding with tiled GEMM execution is essential for high-performance compressed LLM inference.
}
\begin{figure}[t]
    \centering
    \includegraphics[width=\linewidth]{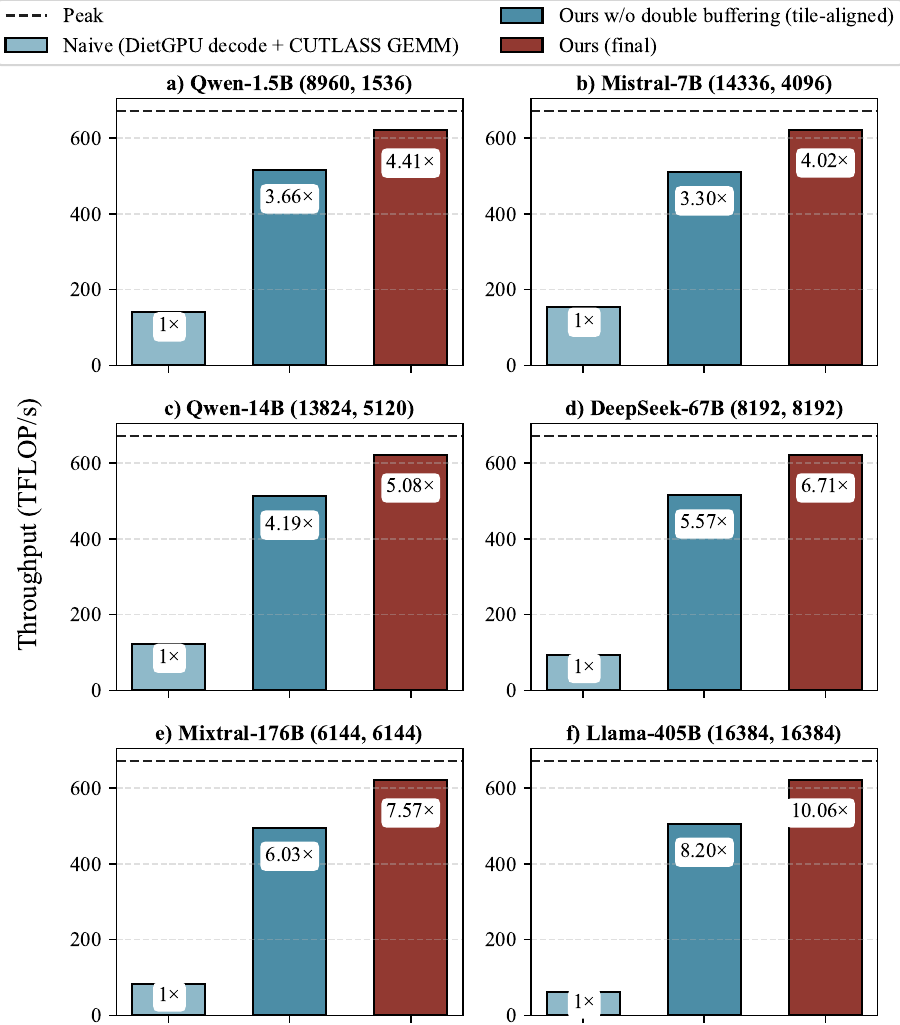}
    \caption{Breakdown of throughput improvements from a naive ANS decoder to the fused decompression GEMM.}
  \label{fig:breakdown}
  \vspace{-10pt}
\end{figure}

\begin{figure*}[t]
    \centering
\begin{minipage}[t]{0.49\linewidth}

    \includegraphics[width=\linewidth]{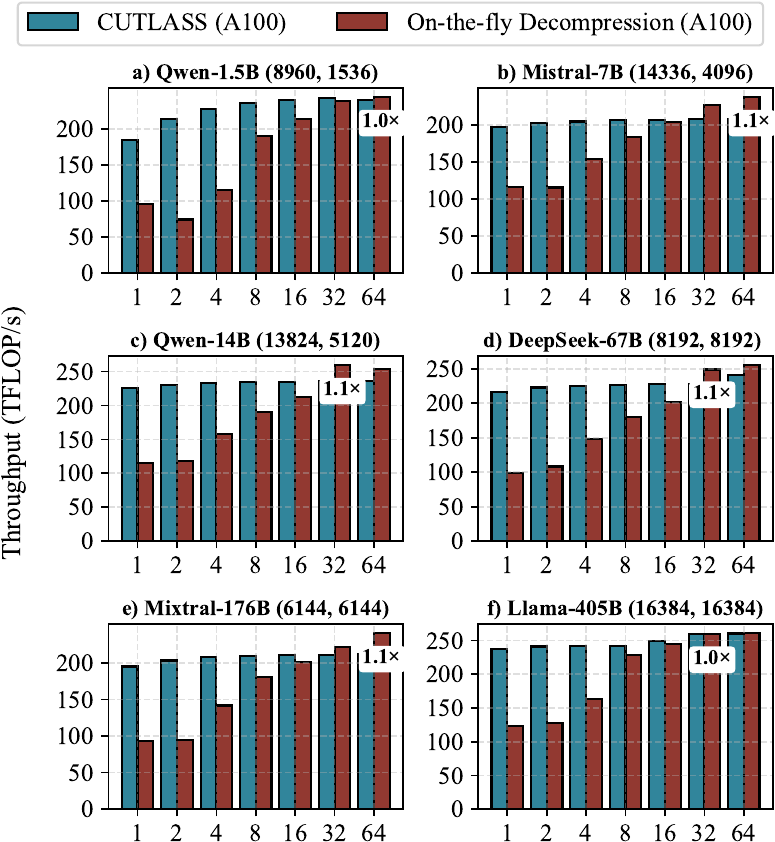}
    \caption{Comparison between our on-the-fly decompression path and the CUTLASS~\cite{cutlass2025}  baseline across varying batch sizes with 4096 input tokens on NVIDIA A100. Weight matrix dimensions are given as $(a,b)$}
    \label{fig:a100_perf}
\end{minipage}
\hfill
\begin{minipage}[t]{0.49\linewidth}
    \centering
    \includegraphics[width=\linewidth]{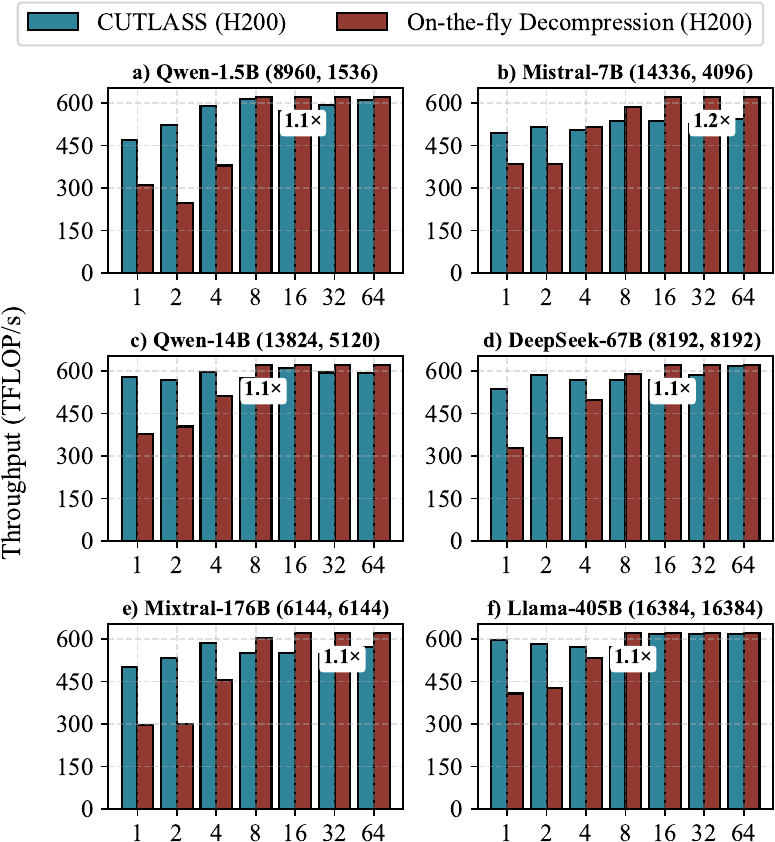}
    \caption{Comparison between our on-the-fly decompression path and the CUTLASS~\cite{cutlass2025}  baseline across varying batch sizes with 4096 input tokens on NVIDIA H200. Weight matrix dimensions are given as $(a,b)$}
      \label{fig:h200_perf}
\end{minipage}

\end{figure*}

\subsection{Performance Analysis}

To evaluate the performance of the on-the-fly decompression GPU kernel, we conducted three complementary experiments that examined the system from different perspectives. First, we present a step-by-step optimization breakdown that quantifies the performance improvement from a naive ANS decoder to our final fused decompression GEMM pipeline. Second, we evaluated kernel-level performance across different GPU architectures (A100 and H200) under a representative inference workload with sequence length 4096. Finally, we compare our approach with KTransformer\cite{10.1145/3731569.3764843}, a system designed to handle out-of-memory (OOM) scenarios by offloading model weights, to demonstrate the end-to-end advantages of the compression-based approach for large-model inference.

\subsubsection{Optimization breakdown of the fused decompression pipeline}
\rv{
\Cref{fig:breakdown} shows the step-by-step throughput improvements from a naive GPU implementation (DietGPU ANS decode + CUTLASS GEMM) to our final fused decompression GEMM kernel.

The first improvement comes from aligning decompression with the GEMM tiling schedule. When decoding and GEMM are executed as separate stages, frequent synchronization and global-memory traffic between the two kernels is needed. Our tile-aligned pipeline eliminates these barriers by decoding weight tiles directly into the layout required by the GEMM microkernel, allowing decompression and computation to proceed in a tightly overlapped fashion. This optimization alone improves throughput by $3.3\!\times$–$8.2\!\times$ across models (e.g., $3.66\!\times$ for Qwen-1.5B and $8.20\!\times$ for Llama-405B).

Despite achieving decompression throughput at the GPU register level in the TB/s range, the decoding stage can still slow down the overall GEMM execution. Therefore, we introduce double buffering in shared memory to prefetch and decode the next tile while the current tile is being consumed by GEMM.
This producer–consumer pipeline hides most of the decompression latency and further increases utilization of tensor cores. With this optimization, the final kernel achieves $4.0\!\times$–$10.1\!\times$ speedup over the naive baseline (e.g., $4.41\!\times$ for Qwen-1.5B, $6.71\!\times$ for DeepSeek-67B, and $10.06\!\times$ for Llama-405B).
The gains become more pronounced for larger matrices, where the decompression overhead can be better amortized and the overlapped pipeline keeps the compute units closer to the peak GEMM throughput.
}
\subsubsection{Cross-GPU kernel performance on A100 and H200}

\Cref{fig:a100_perf} and \Cref{fig:h200_perf} compare our fused on-the-fly decompression kernel with the CUTLASS baseline across varying batch sizes on A100 and H200 GPUs. Across both architectures, throughput increases with batch size as tensor-core utilization improves. As batching grows, our implementation approaches the performance of the native CUTLASS GEMM and in several cases slightly exceeds it.

On A100, our kernel achieves performance close to the baseline across all evaluated models, typically within about $1.0\!\times$–$1.1\!\times$ of CUTLASS at larger batch sizes. This shows that integrating tile-level decompression into the GEMM pipeline introduces minimal overhead while maintaining high tensor-core utilization.

\rv{The advantage becomes more visible on H200. Due to its larger on-chip memory capacity and improved memory subsystem, the tile-level pipeline can overlap decompression and computation more effectively. As a result, our implementation not only matches but occasionally surpasses the CUTLASS baseline, achieving up to about $1.2\!\times$ speedup at larger batch sizes. These results demonstrate that the fused decompression GEMM design scales well across GPU architectures and becomes increasingly effective as the batch size grows.
}

\begin{figure*}[t]
    \centering
\begin{minipage}[t]{0.49\linewidth}
    \includegraphics[width=\linewidth]{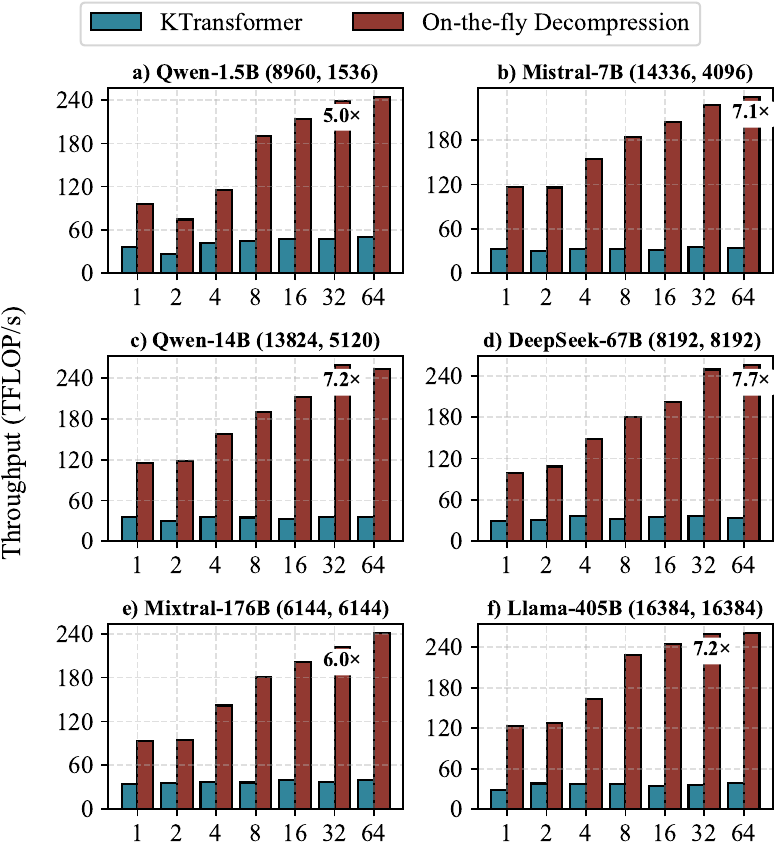}
    \caption{Comparison between our on-the-fly decompression path and the KTransformer~\cite{10.1145/3731569.3764843} baseline across varying batch sizes in the prefill stage with a 4096-token sequence length. Weight matrix dimensions are given as $(a,b)$}
    \label{fig:prefill_kt}
\end{minipage}
\hfill
\begin{minipage}[t]{0.49\linewidth}
    \includegraphics[width=\linewidth]{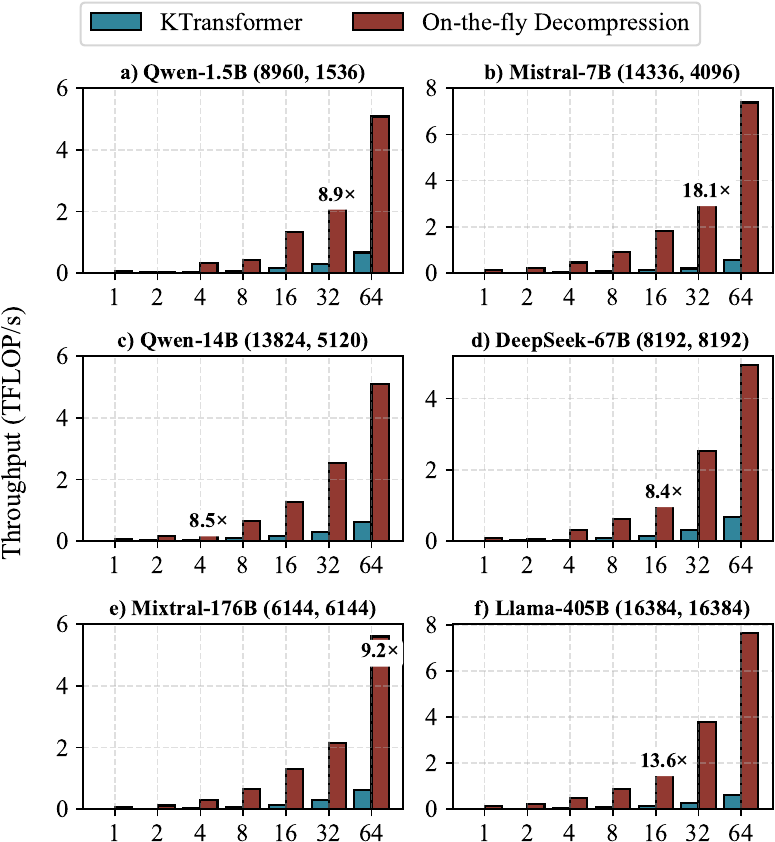}
     \caption{Comparison between our on-the-fly decompression path and the KTransformer~\cite{10.1145/3731569.3764843} baseline across varying batch sizes in the decode stage. Weight matrix dimensions are given as $(a,b)$}
      \label{fig:decode_kt}
\end{minipage}
\end{figure*}

\subsubsection{End-to-End comparison with KTransformer under memory constraints}

\Cref{fig:prefill_kt} reports the prefill-stage throughput of a single representative linear layer drawn from six LLMs, comparing our on-the-fly decompression path against KTransformer across varying batch sizes. In KTransformer, the raw full-precision weight matrix for this layer does not fit into GPU memory, requiring the layer weight to reside in CPU memory and be streamed over PCIe during execution. %

With our lossless tile-level compression, the compressed form of the same layer fits fully in GPU memory. As a result, execution remains entirely on-device, eliminating CPU–GPU streaming and yielding consistently higher throughput, up to $7.7\!\times$ in our layer-level evaluations. This demonstrates that our method provides a practical alternative for improving out-of-memory performance, enabling high throughput without any loss of numerical accuracy.

\Cref{fig:decode_kt} shows that the decode stage exhibits even larger improvements, where up to $18.1\!\times$ throughput improvements are achieved, as weight access becomes the dominant bottleneck when the baseline must stream parameters over PCIe. By eliminating these host–device transfers, our on-the-fly decompression path substantially increases throughput.

\rv{
\subsection{Metadata Overhead Analysis}
\label{sec:metadata-overhead}
We explicitly quantify the metadata footprint of our ANS-compressed tiles. Each compressed tile stores a per-tile offset entry into the compressed buffer, while a single ANS codebook is shared across all tiles in the layer. The compact
representation is therefore
\begin{equation}
\text{Metadata} = \underbrace{4\,\text{B}\cdot N_{\text{tiles}}}_{\text{offset table}} +
\underbrace{(2^{b}\cdot 4\,\text{B})}_{\text{shared codebook}},
\end{equation}
where $b$ is the ANS probability bits (default $b=12$). For fp16 weights, the
tile count is $N_{\text{tiles}}=(K/K_{\text{tile}})\cdot(N/N_{\text{tile}})$ and
we report both percentage of uncompressed size and the effective-bit overhead
(16\,$\times$ metadata/weight).

\paragraph{A100 (32$\times$128)}
The overhead on global memory is only 0.052\%--0.108\% of fp16 weights, i.e., an effective-bit overhead of 0.0083--0.0173 bits/weight. This is sufficiently small that the end-to-end compression rate remains close to the Shannon bound for the payload entropy.

\paragraph{H200 (64$\times$256)}
Since H200 has a larger share memory size compared to A100, it allows large tile size up to 64$\times$256, and the metadata overhead becomes only 0.015\%--0.072\% of uncompressed fp16 weights,
corresponding to 0.0024--0.0115 effective bits per weight. In this regime, the
overhead is further reduced, so the achieved bitrate is very close to the Shannon
limit.
}

\begin{table}[t]
\centering
\caption{Metadata overhead for \texttt{bfloat16} weights (shared codebook + offset table).}
\label{tab:metadata-overhead}
\setlength{\tabcolsep}{3pt}
\begin{tabularx}{\linewidth}{lccc|ccc}
\toprule
\multirow{2}{*}{Model} & \multicolumn{3}{c}{A100 (32$\times$128)} & \multicolumn{3}{c}{H200 (64$\times$256)} \\
\cmidrule(lr){2-4} \cmidrule(lr){5-7}
 & KB & $\%$ of layer & Eff. bits & KB & $\%$ of layer & Eff. bits \\
\midrule
Qwen-1.5B & 29.1 & $0.108\%$ & 0.0173 & 19.3 & $0.072\%$ & 0.0115 \\
Mistral-7B & 72.0 & $0.063\%$ & 0.0100 & 30.0 & $0.026\%$ & 0.0042 \\
Qwen-14B & 83.5 & $0.060\%$ & 0.0097 & 32.9 & $0.024\%$ & 0.0038 \\
DeepSeek-67B & 80.0 & $0.061\%$ & 0.0098 & 32.0 & $0.024\%$ & 0.0039 \\
Mixtral-176B & 52.0 & $0.071\%$ & 0.0113 & 25.0 & $0.034\%$ & 0.0054 \\
Llama-405B & 272.0 & $0.052\%$ & 0.0083 & 80.0 & $0.015\%$ & 0.0024 \\
\bottomrule
\end{tabularx}
\end{table}

\section{Related Work}
\label{sec:related_work}

\noindent\textbf{Data compression on GPU.}
GPU-based compression has been explored extensively in HPC and ML systems to reduce memory traffic and accelerate data movement. nvCOMP~\cite{nvCOMP2025} is the most widely deployed GPU compression library, offering optimized CUDA implementations of LZ4, Snappy, GDeflate, and Bitcomp. However, nvCOMP is closed source and exposes only coarse-grained, host-driven APIs, preventing developers from triggering decompression from inside GPU kernels or at higher granularity. Such lack of device-level control makes it impossible to fuse decompression directly into GEMM pipelines or overlap decoding with computation. Although more entropy-efficient codecs exist, integrating them into LLM inference requires tight alignment with tile-level weight access. Our work addresses this gap via tile-granular ANS decompression tightly coupled with GEMM execution, enabling optimizations that nvCOMP's opaque interface cannot support.

\rv{\noindent\textbf{NVIDIA inline compression} is a hardware mechanism that compresses cache lines in the GPU memory hierarchy to reduce DRAM traffic; it is designed for general workloads without application-level knowledge. In contrast, our work focuses on model-aware compression of LLM weights and targets the inherent redundancy in the weight distribution itself. Rather than relying on opportunistic hardware compression, our approach enables near Shannon's limit compression through distribution-aware encoding integrated with the inference pipeline. As a result, the two techniques operate at different layers of the system stack and are complementary.}

\noindent\textbf{Model compression with GPU codecs.}
LLM.265~\cite{xuLLM265Video2025} repurposes video codecs (H.264/H.265) as tensor compressors for LLMs. While leveraging NVENC/NVDEC hardware is attractive, video engines sustain only $\sim$1.1--1.3~GB/s, two to three orders of magnitude below HBM bandwidth, making them the bottleneck. Furthermore, optimizing for perceptual quality rather than numerical fidelity yields accuracy degradation exceeding 5\% at sub-3-bit. Our method is strictly lossless and designed for GPU-resident LLM inference.

\noindent\textbf{Quantization, pruning, and low-rank compression.}
Extensive work reduces model size using lossy compression techniques, including low-bit quantization~\cite{dettmersQloraEfficientFinetuning2023,heAlphaDecayModulewiseWeight2025,linAwqActivationawareWeight2024,kimOutlierMattersStatistical2025}, group-wise and adaptive quantization~\cite{frantarGptqAccuratePosttraining2022,linAwqActivationawareWeight2024,panSmoothquantAccurateEfficient2023,xiaoSmoothquantAccurateEfficient2023,nagelAdaptiveRoundingPosttraining2020}, pruning~\cite{maLlmprunerStructuralPruning2023,gaoDispllmDimensionindependentStructural2024,yangLacoLargeLanguage2024}, and low-rank decomposition~\cite{huLoraLowrankAdaptation2022,renLowrankPruneandfactorizeLanguage2024,koohpayeganiNolaCompressingLora2023}. While effective, these methods inevitably introduce approximation error or accuracy loss. Our approach is orthogonal: it is fully lossless and can further compress quantized or pruned models. As our entropy analysis shows, even low-bit formats such as FP4, INT4, SmoothQuant, and AWQ retain significant statistical redundancy that lossless compression can exploit.

\section{Conclusion}
\label{sec:conclusion}
We identify a significant gap between the storage bitwidth of LLM weights and their information-theoretic entropy, revealing substantial redundancy even in low-bit formats. To exploit it, we propose a tile-level on-the-fly ANS decompression framework aligned with the GEMM execution pipeline. By decoding weight tiles directly in shared memory and overlapping decompression with tensor-core computation, our fused GPU kernel achieves near-optimal memory reduction without performance overhead, outperforms state-of-the-art lossless methods, and enables larger batch sizes within the memory budget, yielding up to 1.6$\times$ throughput improvement for LLM serving. \emph{Future work} will explore extending on-the-fly entropy coding to the \emph{KV cache} for efficient long-context decoding.

\section*{Acknowledgment}
This research/project is supported by the Ministry of Education AcRF Tier
2 grant (No. MOE-T2EP20224-0020) and Tier 1 grant (No. T1 251RES2315) in Singapore, the Google South \& Southeast Asia Research Award
2025.

\bibliographystyle{IEEEtranS}
\bibliography{refs}

\end{document}